%% file: template.tex
\documentclass{article}
\usepackage{dcase2021,amsmath,graphicx,url,times,booktabs, tabularx}

\usepackage{color, colortbl}
\definecolor{Gray}{gray}{0.9}

\title{A dataset of dynamic reverberant sound scenes with directional interferers for sound event localization and detection}

\name{Archontis Politis$^{1}$,
      Sharath Adavanne$^{1}$,
      Daniel Krause$^{1}$,
      Antoine Deleforge$^{2}$
      }
\secondlinename{	  
      Prerak Srivastava$^{2}$, 
      Tuomas Virtanen$^{1}$,
      }
\address{$^1$ Audio Research Group, Tampere University, Tampere, Finland\\          
        $^2$ Universite de Lorraine, CNRS, Inria, LORIA, F-54000 Nancy, France
 }

\begin{document}

\ninept
\maketitle

\begin{sloppy}

\begin{abstract}
This report presents the dataset and baseline of Task 3 of the DCASE2021 Challenge on Sound Event Localization and Detection (SELD). The dataset is based on emulation of real recordings of static or moving sound events under real conditions of reverberation and ambient noise, using spatial room impulse responses captured in a variety of rooms and delivered in two spatial formats. The acoustical synthesis remains the same as in the previous iteration of the challenge, however the new dataset brings more challenging conditions of polyphony and overlapping instances of the same class. The most important difference of the new dataset is the introduction of directional interferers, meaning sound events that are localized in space but do not belong to the target classes to be detected and are not annotated. Since such interfering events are expected in every real-world scenario of SELD, the new dataset aims to promote systems that deal with this condition effectively. A modified SELDnet baseline employing the recent ACCDOA representation of SELD problems accompanies the dataset and it is shown to outperform the previous one. The new dataset is shown to be significantly more challenging for both baselines according to all considered metrics. To investigate the individual and combined effects of ambient noise, interferers, and reverberation, we study the performance of the baseline on different versions of the dataset excluding or including combinations of these factors. The results indicate that by far the most detrimental effects are caused by directional interferers.
\end{abstract}

\begin{keywords}
Sound event localization and detection, sound source localization, acoustic scene analysis, microphone arrays
\end{keywords}

\section{Introduction}
\label{sec:intro}

Sound event localization and detection (SELD) is an audio processing task that aims to jointly detect temporally target classes of sound events and localize them in space when active. In that sense it differs from the classic sensor array task of sound source localization (SSL) which utilizes only spatial information to detect, localize, and track sources independently from their signal content \cite{evers2020locata}. It also differs from the popular sound event detection (SED) task which is focused on the temporal detection and classification part, omitting the spatial information of the scene. The spatiotemporal characterization of the scene produced by SELD makes it suitable for a range of applications such as robot audition and machine listening in general \cite{valin2003robust, He2018}, acoustic monitoring \cite{valenzise2007scream, do2018rish}, smart home environments \cite{do2018rish,brdiczka2008learning}, improved human-machine interaction \cite{cech2013active}, speech recognition \cite{he2018deep}, and sonic information visualization \cite{matsinos2008spatio}, among others.

Research interest in SELD grew quickly during the last couple of years, with deep learning methods handling the task jointly \cite{adavanne2018sound}, or fusing information from solving individual subtasks of SED and SSL \cite{Cao2019, nguyen2020sequence}. This interest culminated in the task becoming part of the DCASE Challenge
in 2019, with participants bringing novel approaches to the problem, summarized in \cite{politis2020overview}. The dataset used in the challenge \cite{Politis2020} included sound scenes from two different array formats with sound events spatialized in both azimuth and elevation using spatial room impulse responses (SRIRs) of real rooms. Additionally, spatial ambient noise captured in situ was added to the recordings. For the next iteration of the task in the DCASE Challenge 2020,
 a new dataset was generated based on SRIRs from additional rooms with more realistic and challenging conditions beyond the limitations of the first one \cite{Politis2020}. More specifically, the discrete grid of potential directions-of-arrival (DOAs) of the older dataset was replaced with continuous DOA trajectories and, apart from static events, moving sources using interpolated SRIRs were emulated at different speeds. Furthermore, the newer SRIRs were captured in rooms of more diverse acoustical properties and from a wider range of distances, resulting in longer reverberation times and more challenging direct-to-reverberant ratios (DRRs).

The second iteration of the SELD task in DCASE2020 brought additional innovations, with participants experimenting with homogeneous joint loss functions \cite{shimada2021accdoa, Phan2020}, self-attention layers \cite{Phan2020, wang2021four}, advanced spatial augmentation strategies \cite{shimada2021accdoa, wang2021four}, combinations of model-based localization with learning-based SED \cite{Nguyen2020c, Perez-Lopez2020}, data-based fusion of individual SSL and SED systems \cite{Nguyen2020c, nguyen2021general}, and event- or track-based prediction modeling, instead of class-based prediction \cite{Cao2020, Perez-Lopez2020}. The latter development specifically tried to address the case of same-class events occurring simultaneously \cite{nguyen2020sequence, Cao2020, cao2021improved}, a case that distinguishes the SELD task from SED and becomes possible mainly due to spatial information. Research following the DCASE2020 challenge investigated fusion of pre-trained SED and SSL models \cite{nguyen2021general}, or parameter sharing between joint, semi-joint SELD models, and models fusing SSL and SED subsystems \cite{cao2021improved}.

This report introduces the new \textbf{TAU-NIGENS Spatial Sound Events 2021}\footnote{\url{https://doi.org/10.5281/zenodo.4844825}} dataset and the baseline\footnote{\url{https://github.com/sharathadavanne/seld-dcase2021}} of the SELD challenge task in DCASE2021\footnote{\url{http://dcase.community/challenge2021/task-sound-event-localization-and-detection}}. The major difference of this dataset with the previous one is the introduction of localized interfering events outside of the target classes. This condition, naturally encountered in a real environment, introduces new challenges to the task. Apart from the dataset and baseline description, we present an extensive evaluation of the baseline on different versions of the dataset with and without the presence of ambient noise, directional interferers, and reverberation.

\section{Dataset}
\label{sec:dataset}

Similarly to the dataset of the previous iteration, the current one consists of 800 one-minute spatial recordings, of which 600 constitute the development set of the dataset, and the other 200 the evaluation set. The recordings are sampled at 24kHz, and they are offered in two 4-channel spatial audio formats, the raw signals of a tetrahedral microphone array and first-order Ambisonics, abbreviated as MIC and FOA for the rest of the paper. Detailed descriptions of the formats in terms of their directional encoding properties can be found in the previous challenge dataset reports \cite{Adavanne2019a, Politis2020}.

\subsection{Sound events}

The sound event samples are sourced from the \emph{NIGENS general sound events database} \cite{trowitzsch2019nigens}, which consists of 14 classes of specific sound types, and an additional general one with disparate sounds not belonging to any of the other classes. We use the sounds in the 12 classes \emph{alarm, crying baby, crash, barking dog, female scream, female speech, footsteps, knocking on door, male scream, male speech, ringing, phone, piano} as target events, and the sounds in the classes \emph{running engine, burning fire} and the general class as directional interferers. This division results in about 500 distinct sound samples distributed across the target events of the dataset, and about 400 across the interfering events.

\subsection{Dataset synthesis}

The synthesis of the spatial sound reccordings are based on a collection of SRIRs acquired continuously along measurement trajectories inside 13 enclosures of Tampere University. The RIR collection and synthesis process is described in more detail in \cite{Politis2020}. We summarize briefly the acoustical properties of the dataset. SRIRs are extracted along the measurement trajectories with an approximate resolution of 1 degree, resulting on about 1184 to 6480 possible RIRs/DOAs per room, depending on the type (circular/linear) and number of measurement trajectories. Events added in a single recording can be static or moving. The source position for a static event is drawn randomly from the pool of SRIRs of a single room used in that recording, while moving events are synthesized for one of the measured trajectories in the room. Moving events are synthesized to have an approximate speed of 10$^\circ$/sec, 20$^\circ$/sec, or 40$^\circ$/sec, drawn randomly.  The dataset is split into 8 folds with distinct rooms and samples in each of them. Distinct rooms result in different reverberation conditions, and even though similar ranges of DOAs may occur between rooms, the source distance, DRR, and reverberation conditions are distinct between folds for a certain DOA.

The events are laid out in layers in each recording, with the total number of layers determining the maximum polyphony possible. The parameter determining the density of events per layer and, hence, the average per-frame polyphony is the total gap time distributed between events in each layer. A larger gap time results in fewer events per layer and a lower average polyphony, while a smaller gap time results in higher event density and average polyphony. The last event per layer is truncated to fit the total 1 minute duration. For the present dataset there are three layers of target events and an additional layer of interfering events, resulting in a total maximum polyphony of 4. In addition to the spatialized reverberant events, multichannel ambient noise that was collected in each room with the same recording setup as the SRIRs is truncated to 1 minute segments and added to the event mixtures. The noise is scaled to result in signal-to-noise ratios (SNRs) ranging from noiseless (30dB) to noisy (6dB) conditions, with respect to the total energy of the target events in the recording excluding silences. A depiction of the layering of events in one recording is shown in Fig.~\ref{fig:recording}.

\begin{figure}[t]
  \centering
  \centerline{\includegraphics[width=0.4\textwidth ,keepaspectratio]{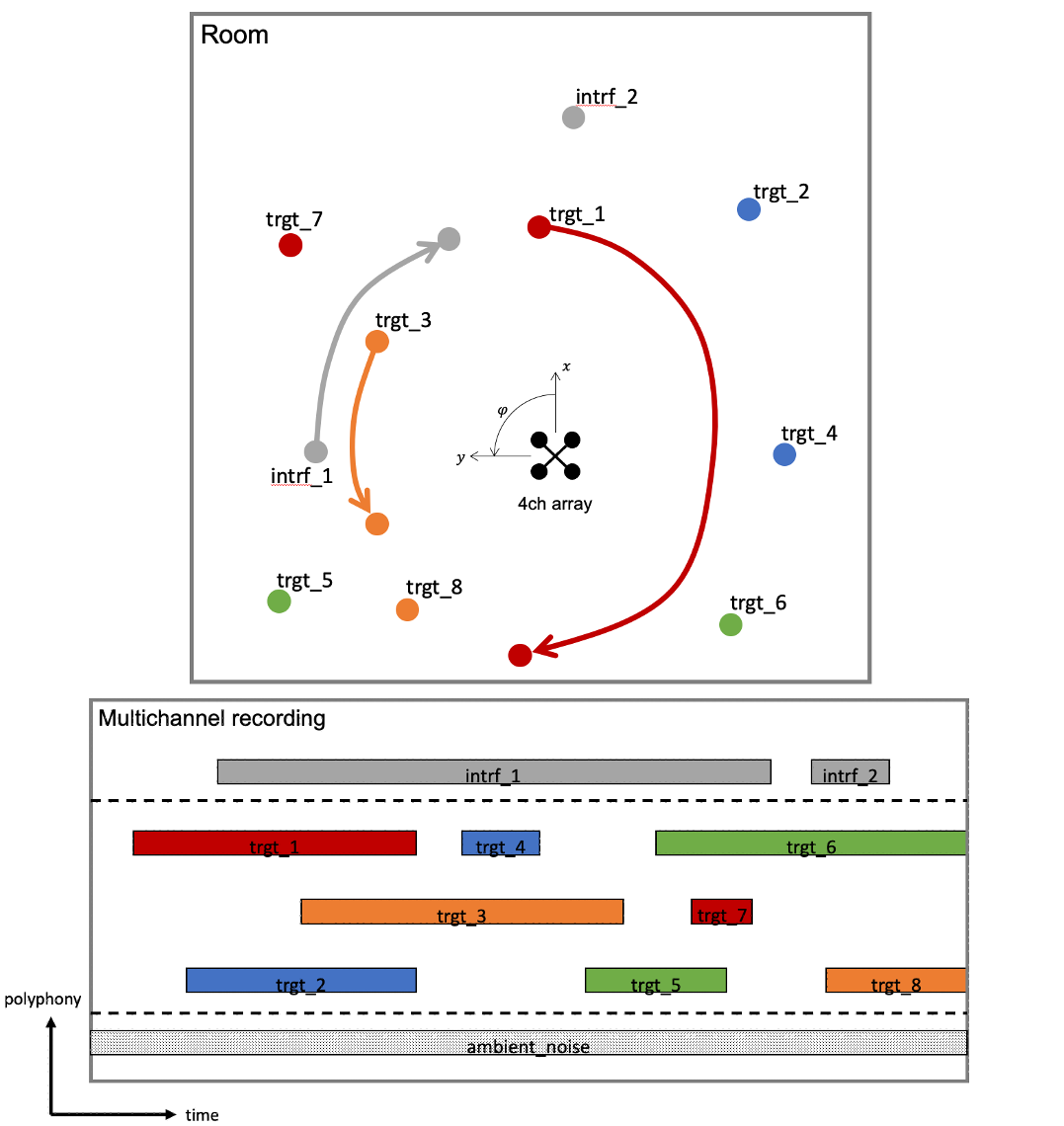}} 
  \caption{A graphic depiction of an emulated recording in the dataset, with colored objects indicating target classes, gray objects indicating interferers and ambient noise, and arrows indicating moving events.}
  \label{fig:recording}
\end{figure}

\subsection{Differences with DCASE2020 task 3 dataset}

Even though the acoustical and synthesis characteristics of the new dataset are similar to the dataset of the previous DCASE2020 Challenge, the following differences make it more challenging:
\begin{enumerate}
    \item Directional interferers, out of the target classes of a detection system, are common in real conditions and they add to the challenge by forcing a strong joint modeling and training strategy that can learn to ignore them.
    \item The overall maximum polyphony is increased from 2 to 3 target events.
    \item The recordings are not anymore divided into recordings with no overlap (polyphony 1), and recordings with two simultaneous events (polyphony 2). Instead all recordings have the maximum level of polyphony, with all intermediate levels (from silence to 3 simultaneous target events + interference) varying during the duration of the recording. This choice reflects more natural recording conditions in a real dataset.
    \item Even though the dataset of DCASE2020 had instances of the same class occurring at the same time, such occurrences were fairly rare. In the present dataset, these occurences have been increased in order to give a clear advantage to systems that can resolve this difficult but realistic case.
\end{enumerate}

\section{Baseline}
\label{sec:baseline}

\begin{figure}[t]
  \centering
  \centerline{\includegraphics[width=0.49\textwidth ,keepaspectratio]{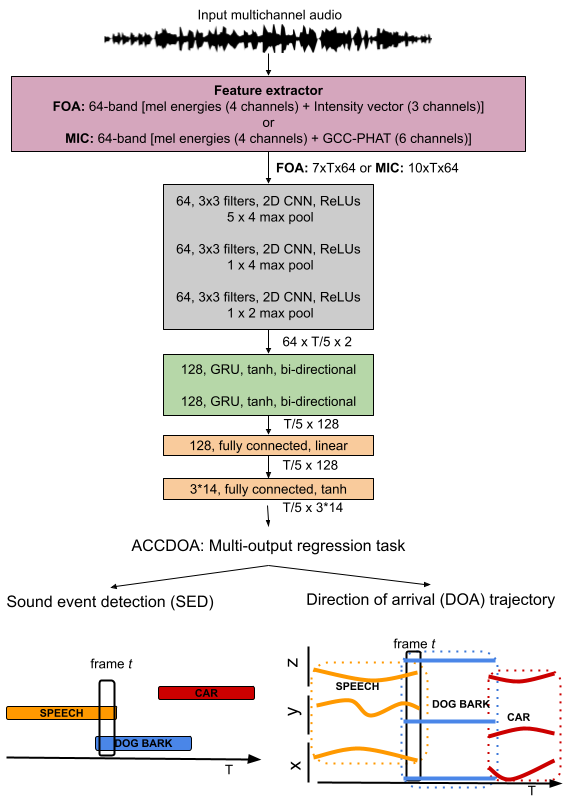}} 
  \caption{Convolutional recurrent neural network with ACCDOA loss for SELD.}
  \label{fig:crnn}
\end{figure}

Similar to the previous iterations of the challenge, we adopt a modified version of SELDnet \cite{adavanne2018sound} as the baseline method, due to its conceptual simplicity. Its architecture remains a convolutional recurrent neural network (CRNN) receiving multichannel log-mel spectrograms as inputs, together with acoustic intensity vectors \cite{pulkki2017first} for the FOA dataset, and generalized cross-correlation (GCC-PHAT) sequences for the MIC dataset, added as extra channels. The baseline implementation extracts log-mel spectrograms in  64 mel-bands from 1024-point FFTs, using a 40 ms window and 20 ms hop length at 24kHz. The intensity vectors are similarly extracted for every FFT bin and aggregated in the same number of mel-bands as the spectrograms, while the GCC sequences are also truncated to the same number of lag values as the mel-bands, adopted from \cite{Cao2019}. More details on the architecture and features can be found in \cite{Politis2020}.

The only difference of the current SELDnet baseline with respect to the previous DCASE challenge iteration is the output format and the respective loss function. The original SELDnet architecture employs separate output branches for detection and localization, with as many classification outputs and as many localization regressors as the number of classes. In the current baseline, we adopt the \emph{activity-coupled cartesian direction of arrival} representation (ACCDOA) introduced in DCASE2020 Challenge by Shimada et al. \cite{shimada2021accdoa}, which unifies the SED and SSL losses into a single homogeneous regression loss, simplifying the overall architecture by removing the detection branch, while simultaneously improving its performance. Using the ACCDOA representation, the network receives a sequence of $T$ STFT frames of multichannel features and outputs $T/5\times 3$ Cartesian vector coordinates for each of the target classes, with the direction of each vector indicating DOA and the vector length indicating class activity probability. The reduction in temporal resolution is intended to match the 100 ms resolution of annotations in the challenge. A block diagram of the current baseline architecture is shown in Fig.~\ref{fig:crnn}.

\section{Evaluation}
\label{sec:evaluation}
The dataset and baseline are delivered to the challenge participants at the commencement of the challenge, along with the development set of the dataset consisting of the 6 first folds, while the last two folds are made available during the evaluation phase of the challenge. Participants are required to report results on the test set of the development set using the predefined split of Table~\ref{tab:splits}, so that conclusions can be drawn among the submissions on the same configuration. However, during the evaluation phase, participants have to report results on the evaluation dataset only, using the development dataset for training and validation in any way they see fit.

\begin{table}[h]
\centering  
\renewcommand\thetable{1}
\caption{Evaluation setup}
\begin{tabular}{l|ccc}
 & \multicolumn{3}{c}{\textbf{Splits}} \\ \cline{2-4}
\textbf{Dataset} & \textbf{Training} & \textbf{Validation} & \textbf{Testing} \\ \hline
\textbf{Development}  & 1,2,3,4 & 5 & 6 \\
\end{tabular}
\label{tab:splits}
\end{table}

The submissions are evaluated using the same combination of joint detection/localization metrics studied in \cite{mesaros2019joint, politis2020overview} and introduced in DCASE2020. Closer to SED evaluation, the localization-dependent error rate ($ER_X$) and F1-score ($F_X$) express detection performance but they penalize correct detections that occur further from the reference than some threshold distance $X$. On the other hand, the class-dependent localization error ($LE_\mathrm{CD}$) and localization recall ($LR_\mathrm{CD}$) are inspired by classical localization metrics, but are computed for each class individually before being averaged. The $LE_\mathrm{CD}$ is a mean angular localization error after pairing the predicted DOAs to their closest reference DOAs, while $LR_\mathrm{CD}$ is a simple recall metric on the detected localized events without any spatial threshold. Since in the SELD case there can be multiple simultaneous references of the same class, the detection metrics are modified to consider multiple instances of the same class and penalize cases where, e.g., only one of the predictions belong to that class. For the exact formulation of the metrics the reader is referred to \cite{politis2020overview}. The submissions are first ranked for each of the four metrics individually, and the final rank of each system is determined by the sum of the four individual ranks.

\input{table_baselines}

\section{Results}
\label{sec:Results}

In order to evaluate the performance of the new baseline utilizing the ACCDOA loss, we compare it against the previous SELDnet baseline of DCASE2020,
on the development set of DCASE2020 and the current one. Table~\ref{tab:baseline-results} shows a clear improvement of the ACCDOA version in all metrics. Especially in the more challenging new dataset, the ACCDOA loss brings large gains in detection and improves localization accuracy by about 25\%. A significant decrease of performance for both methods is also observed from the DCASE2020 dataset to the DCASE2021 dataset. This suggests that the new dataset is more challenging, as intended.

To get a more detailed picture on the effect of the various components in the scene, namely reverberation, ambient noise, and directional interferers, we generate various versions of the dataset including those components in various combinations. More specifically, the \emph{targets}, \emph{targets+ambience}, \emph{targets+interferers}, \emph{targets+ambience+interferers} develop from the presence of targets only, to the inclusion of ambient noise or interferers separately, to the full dataset combining all components. Excluding the effect of reverberation is less straightforward due to the use of real SRIRs for the synthesis. In order to generate reverberation-free versions of the dataset, the sound events for each recording in the original dataset are spatialized with anechoic IRs of the same Eigenmike spherical microphone array used to capture the SRIRs. The anechoic array IRs are computed for the same measurement trajectories and DOAs as the measured SRIRs in each room, and stored in a similar data structure. Additionally, each IR is delayed and scaled according to the source distance of the respective measured SRIR, following an inverse distance law and a speed of sound of $c=343 m/sec$. Delaying and scaling ensures that the events between the reverberant and non-reverberant versions are approximately time-aligned and with comparable distance-dependent attenuation. The Eigenmike responses were measured in an equirectangular grid of $5^{\circ}$ azimuth and $5^{\circ}$ elevation in the large anechoic chamber of Aalto University, as described in \cite{tervo2015direction}. Since the DOAs in the reverberant dataset do not necessarily coincide with the measurement grid of the array, array response interpolation is performed to recover anechoic IRs at the DOAs of the measured SRIRs, based on a spherical harmonic expansion of the array steering vectors, as in \cite{politis2017comparing}.

The results are presented in Table~\ref{tab:baseline-results}. As expected, reverberation affects negatively all combinations, increasing error rates and decreasing F-scores and localization recall in a consistent manner between the same scenarios. Additionally, it decreases localization accuracy by $2^{\circ}$--$4^{\circ}$. Inclusion of the ambient noise has a small but noticeable effect when added to the targets, without interferers. The small effect may be due to the large range of possible positive SNRs (6--30dB) distributed uniformly across the recordings, with most of them having adequate SNR to be unaffected by the noise presence. Interestingly, together with directional interference, inclusion of ambient noise seems to even improve certain results slightly.  This may be due to potential regularization effects of noise and is worth further investigation.

The most detrimental effects happen with the inclusion of the directional interferers, proving that this challenging case will need to be taken into account for future SELD systems. Error rate $ER$ increases up to about 40\% in the non reverberant case for the FOA recordings, and up to about 33\% for the MIC recordings. Similarly, in reverberant scenarios, the $ER$ increases up to about 28\% for both FOA and MIC formats. F-scores decrease by up to 40\% on the FOA dataset and up to 50\% on the MIC dataset, for both anechoic and reverberant conditions. The localization recall ($LR$) also drops by about 30\% for both formats and both anechoic and reverberant conditions. Finally, localization errors increase by up to about $7^{\circ}$ in the case of FOA recordings and up to $10^{\circ}$ in the case of MIC recordings, for both anechoic and reverberant conditions. In general, the MIC dataset exhibits a worse performance than FOA in all cases. This fact may be attributed to the input features employed in the baseline for each format. GCC sequences for the MIC format may become very noisy in complex scenes with multiple simultaneous events, while the intensity vectors of the FOA format can potentially retain robustness due to their narrowband nature and sparsity of the event signals in the time-frequency domain.

\input{table_results}

\section{Conclusions}
\label{sec:conclusions}

In this report we describe the new dataset and baseline for the SELD task of the DCASE2021 challenge. The differences with the dataset of the previous iteration of the challenge are highlighted; namely, inclusion of directional interferers, higher polyphony, and higher number of multiple simultaneous same-class event occurences. The evaluation task setup is also described, including a predefined fixed split on the development data for straightforward comparison of the submissions. The new baseline adopts the recent ACCDOA SELD representation introduced in the previous challenge to improve its performance, and is evaluated in the testing split of the development dataset. The new dataset is shown to be significantly more challenging for both the previous and the new baseline according to all considered metrics. A detailed analysis of the new baseline on different versions of the dataset shows that between reverberation, ambient noise, and directional interferers, the latter has the most detrimental effect of the three by far, in all evaluation metrics. 

\newpage
\bibliographystyle{IEEEtran}
\bibliography{refs}

%
%
%
%
%
%
%
%
%

\end{sloppy}
\end{document}

%% file: table_baselines.tex
\begin{table*}[h]
\caption{Comparison between the DCASE2020 baseline and the current one on the development set of DCASE2020 and the current development set. \emph{2020-multi} refers to the previous baseline with separate output branches and losses for detection and localization, while \emph{2021-accdoa} refers to the current baseline with the unified ACCDOA loss.}
\label{tab:baseline-comparison}
\centering
\begin{tabular}{lllllllll}
& \multicolumn{4}{c|}{\textbf{FOA}}                                      & \multicolumn{4}{c}{\textbf{MIC}} \\ \cline{2-9} 
\multicolumn{1}{l|}{}                                                                                                     & \textbf{$ER_{20^\circ{}}\downarrow$} & \textbf{$F_{20^\circ{}}\uparrow$} & \textbf{$LE_{CD}\downarrow$}   & \multicolumn{1}{l|}{\textbf{$LR_{CD}\uparrow$}}                      & \textbf{$ER_{20^\circ{}}\downarrow$} & \textbf{$F_{20^\circ{}}\uparrow$} & \textbf{$LE_{CD}\downarrow$}   & \multicolumn{1}{l}{\textbf{$LR_{CD}\uparrow$}}    \\ \hline

\multicolumn{9}{l}{\textbf{DCASE2020 development set}} \\
\hline
\multicolumn{1}{l|}{\textbf{2020-multi}}             
&  0.70 &  44.4\% &  24.3$^\circ$ & \multicolumn{1}{l|}{61.9\%}
&  0.71  &  40.4\%  &  25.4$^\circ$  & 55.4\%  \\ \cline{1-1}
\multicolumn{1}{l|}{\textbf{2021-accdoa}}                                        & 0.60 & 51.9\% & 17.9$^\circ$ & \multicolumn{1}{l|}{59.8\%}                         &  0.61  &  48.5\%  &  19.3$^\circ$  & 55.2\%  \\  
\hline

\multicolumn{9}{l}{\textbf{DCASE2021 development set}}    \\ \hline
\multicolumn{1}{l|}{\textbf{2020-multi}}                  & 0.77 & 24.7\% & 32.1$^\circ$ & \multicolumn{1}{l|}{44.8\%}                         & 0.81   & 19.1\%   & 41.6$^\circ$   & 47.4\%  \\ \cline{1-1}
\rowcolor{Gray}
\multicolumn{1}{l|}{\textbf{2021-accdoa}} & 0.73 & 30.7\% & 24.5$^\circ$ & \multicolumn{1}{l|}{40.5\%} & 0.75   & 23.4\%   & 30.6$^\circ$   & 37.8\% 
\end{tabular}
\end{table*}

%% file: table_results.tex
\begin{table*}[t]
\caption{Performance of the DCASE2021 baseline for different versions of the dataset with increasingly adverse conditions. The highlighted row corresponds to the version of the dataset used in the challenge.}
\label{tab:baseline-results}
\centering
\begin{tabular}{lllllllll}
                                                                                                                                 & \multicolumn{8}{c}{\textbf{Development set}}                                                              \\ \cline{2-9} 
                                                                                                                                 & \multicolumn{4}{c|}{\textbf{FOA}}                                      & \multicolumn{4}{c}{\textbf{MIC}} \\ \cline{2-9} 
\multicolumn{1}{l|}{}                                                                                                            & \textbf{$ER_{20^\circ{}}\downarrow$} & \textbf{$F_{20^\circ{}}\uparrow$} & \textbf{$LE_{CD}\downarrow$}   & \multicolumn{1}{l|}{\textbf{$LR_{CD}\uparrow$}}                           & \textbf{$ER_{20^\circ{}}\downarrow$} & \textbf{$F_{20^\circ{}}\uparrow$} & \textbf{$LE_{CD}\downarrow$}   & \multicolumn{1}{l}{\textbf{$LR_{CD}\uparrow$}}    \\ \hline
\multicolumn{9}{l}{\textbf{Non-reverberant results}}                                                                                                                                                                                         \\ \hline
\multicolumn{1}{l|}{\textbf{targets}}                                                                                            & 0.49 & 62.0 & 16.3 & \multicolumn{1}{l|}{65.7}                         & 0.54   & 55.4   & 20.8   & 63.7  \\ \cline{1-1}
\multicolumn{1}{l|}{\textbf{\begin{tabular}[c]{@{}l@{}}targets+\\ ambience\end{tabular}}}                                        & 0.49 & 61.2 & 16.4 & \multicolumn{1}{l|}{65.6}                         & 0.57   & 51.2   & 20.8   & 58.9  \\ \cline{1-1}
\multicolumn{1}{l|}{\textbf{\begin{tabular}[c]{@{}l@{}}targets+\\ interferers\end{tabular}}}                                     & 0.69 & 36.9 & 24.1 & \multicolumn{1}{l|}{45.2}                         & 0.72   & 27.7   & 30.5   & 42.2  \\ \cline{1-1}
\multicolumn{1}{l|}{\textbf{\begin{tabular}[c]{@{}l@{}}targets+\\ ambience+\\ interferers\end{tabular}}}                         & 0.66 & 40.3 & 22.7 & \multicolumn{1}{l|}{46.9}                         & 0.73   & 26.7   & 30.4   & 42.5  \\ \hline
\multicolumn{9}{l}{\textbf{Reverberant results}}                                                                                                                                                                                             \\ \hline
\multicolumn{1}{l|}{\textbf{targets}}                                                                                            & 0.55 & 53.7 & 19.9 & \multicolumn{1}{l|}{61.3}                         & 0.59   & 47.0   & 22.0   & 57.3  \\ \cline{1-1}
\multicolumn{1}{l|}{\textbf{\begin{tabular}[c]{@{}l@{}}targets+\\ ambience\end{tabular}}}                                        & 0.57 & 50.3 & 20.2 & \multicolumn{1}{l|}{59.3}                         & 0.62   & 44.2   & 22.8   & 53.6  \\ \cline{1-1}
\multicolumn{1}{l|}{\textbf{\begin{tabular}[c]{@{}l@{}}targets+\\ interferers\end{tabular}}}                                     & 0.71 & 32.7 & 26.7 & \multicolumn{1}{l|}{44.2}                         & 0.76   & 24.0   & 32.6   & 39.4  \\ \cline{1-1}
\rowcolor{Gray}
\multicolumn{1}{l|}{\textbf{\begin{tabular}[c]{@{}l@{}}targets+\\ ambience+\\ interferers\end{tabular}}} & 0.73 & 30.7 & 24.5 & \multicolumn{1}{l|}{40.5} & 0.75   & 23.4   & 30.6   & 37.8 
\end{tabular}
\end{table*}